\documentclass[aps,prl,twocolumn,showpacs,groupedaddress]{revtex4}

\usepackage{amssymb}
\usepackage{amsmath}
\usepackage{dcolumn}
\usepackage{bm}
\usepackage{graphicx}

\newcommand{\bra}[1]{\langle #1|}
\newcommand{\ket}[1]{|#1\rangle}

\begin{document}

\title{Molecular States in Carbon Nanotube Double Quantum Dots}

\author{M. R. Gr{\"a}ber}
\author{W. A. Coish}

\author{C. Hoffmann$^{\S}$}
\author{M. Weiss}
\author{J. Furer}
\author{S. Oberholzer}
\author{D. Loss}
\author{C. Sch{\"o}nenberger}

\email{christian.schoenenberger@unibas.ch} \affiliation{Institut
f{\"u}r Physik, Universit{\"a}t Basel, Klingelbergstr.~82, CH-4056
Basel, Switzerland }

\date{\today}

\begin{abstract}
We report electrical transport measurements through a
semiconducting single-walled carbon nanotube (SWNT) with three
additional top-gates. At low temperatures the system acts as a
double quantum dot with large inter-dot tunnel coupling allowing
for the observation of tunnel-coupled molecular states extending
over the whole double-dot system. We precisely extract the tunnel
coupling and identify the molecular states by the
sequential-tunneling line shape of the resonances in differential
conductance.

\end{abstract}


\maketitle



The interference of quantum states is one of the most striking
features of nature enabling the formation of molecular bonds. This
bond formation can be studied in coupled quantum dots (artificial
molecules) in regimes that are not accessible in true
molecules~\cite{Livermore,Blick,Holleitner,Petta,Elzermann}.
Additionally, these engineered artificial molecules have been
proposed as logic elements for future applications in spin-based
quantum computing~\cite{Burkard}. Whereas most electrical
transport experiments on coupled quantum dots so far have
investigated GaAs-based semiconductor quantum dots (see
\cite{VanderWiel} and references therein), only recently such
structures have been realized in carbon nanotubes and
semiconducting nanowires~\cite{Mason,Fasth}. These materials are
attractive not just for the relative ease in production, but also
for the fact that superconducting and ferromagnetic contacts have
been demonstrated \cite{Sahoo,Buitelaar,Francheschi}, opening up a
road for various kinds of novel quantum devices \cite{Choi}. In
addition, large spin dephasing times are expected for carbon-based
quantum dots, since the nuclear spin of the dominant isotope
$^{12}C$ is zero, yielding a strongly reduced hyperfine
interaction.

In this letter, we report electrical transport measurements
through a semiconducting single-walled carbon nanotube (SWNT) with
source and drain electrodes and three additional top-gates. In
specific gate-voltage ranges the system acts as a double quantum
dot with {\em large} inter-dot tunnel coupling $t$, allowing for
the observation of a quantum-mechanical superposition of
$\ket{n,m+1}$ and $\ket{n+1,m}$ states where $n$ and $m$ denote
the number of charges on the left and right dot, respectively.
Using an effective single-particle picture, we precisely determine
the tunnel coupling and identify molecular-like states with wave
functions extending over the whole nanotube double dot.

\begin{figure}[b]
\scalebox{0.9}{\includegraphics*[width=\linewidth]{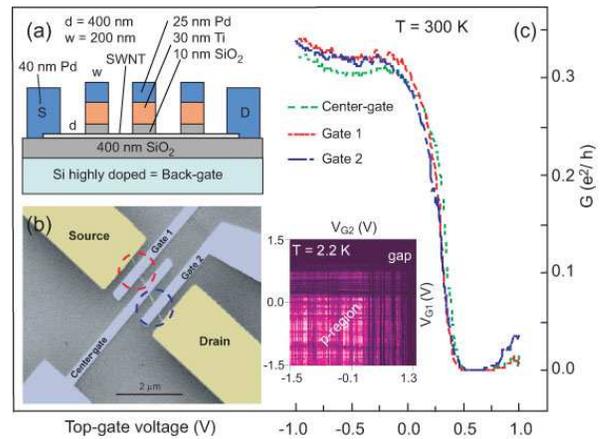}}
\caption{\label{figure1} (a) Schematic of the fabricated device,
with three top-gates as labelled in (b). (b) Scanning electron
micrograph of a sample fabricated identically to the one measured.
The distance from source to drain is \mbox{$2.2$\,$\mu$m}. Dashed
circles denote the regions affected by gates~1~and~2. (c)
Conductance $G$ through the device at \mbox{$T=300$\,K} versus
top-gate voltage. All gates not swept are connected to ground.
Note: Differences between the individual gate scans at
\mbox{$0$\,V} arise from slightly hysteretic gate responses.
Inset: Colorscale plot of $G$ versus gate~1 and gate~2 for fixed
\mbox{$V_{C}=-1$\,V} at \mbox{$2.2$\,K}. Bright corresponds to
\mbox{$0.4$\,e$^2$/h}, dark to \mbox{$0$\,e$^2$/ h}. }
\end{figure}

Single-walled carbon nanotubes were grown by means of chemical
vapor deposition (for details see Ref.~\cite{Juerg}) on a
highly-doped Si substrate covered by an insulating layer of
\mbox{$400$\,nm} SiO$_{2}$. Single nanotubes were selected using a
scanning electron microscope. Three 200 nm wide local gates
equally spaced by \mbox{$400$\,nm} were then defined by means of
standard electron beam lithography and e-gun evaporation of
SiO$_2$, Ti and Pd. Finally, Pd source and drain contacts were
fabricated. Figure~\ref{figure1}(a) shows a schematic of the
device, the materials used, and corresponding film thicknesses. A
scanning electron micrograph of a device
is shown in Fig.~\ref{figure1}(b).

 Room temperature characterization identifies the semiconducting
nature and an intrinsic p-doping state of the nanotube.
Figure~\ref{figure1}(c) shows the linear conductance through the
device as a function of the three top-gate voltages. At a top-gate
voltage of roughly 0.4~V conductance is suppressed indicating that
the chemical potential is shifted into the semiconducting gap of
the tube.
Five identically-prepared devices were tested at room temperature
and showed the same behavior.

Low-temperature measurements were performed in a $^{3}$He cryostat
with a base temperature of \mbox{$290$\,mK}. Differential
conductance $dI/dV_{sd}$ was measured using standard lock-in
techniques with an excitation voltage of typically
\mbox{$7.5$\,$\mu$V} at a frequency of \mbox{$327.7$\,Hz} and an
I/V converter with a gain of \mbox{$10^{7}$\,V/A}. The inset of
Fig.~\ref{figure1}(c) shows a colorscale plot of the linear
conductance versus voltages applied at gates~1~and~2 for a
constant center gate voltage $V_{C}=-1$~V at \mbox{$2.2$\,K}.
Again, applying positive voltages of the order \mbox{$1$\,V} to
any of the top-gates locally shifts the chemical potential into
the energy gap of the intrinsically p-doped SWNT and thus
suppresses electrical transport. Additionally, sweeping gate~1 and
gate~2 leads to pronounced oscillations of the conductance due to
single-electron charging and finite-size effects of the nanotube,
which are accessible at low temperatures. For the measurements
presented in the following, the center and back-gate were kept at
constant voltages \mbox{$V_{C}=-0.1$\,V}, \mbox{$V_{BG}=0$\,V},
respectively, and no magnetic field was applied.

\begin{figure}[th]
\scalebox{0.9}{\includegraphics*[width=
\linewidth]{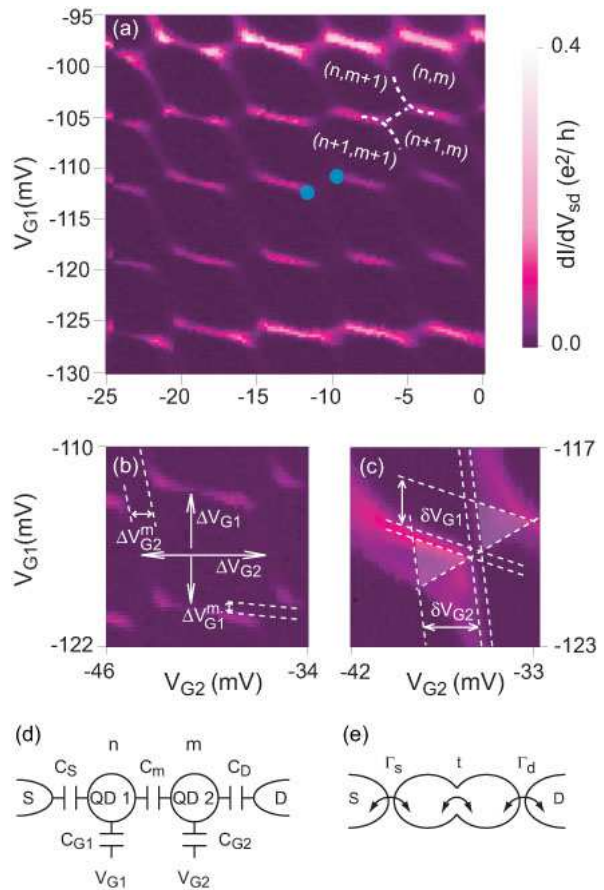}} \caption{\label{figure2} (a)
Colorscale plot of the conductance versus voltage applied on
gate~1 ($V_{G1}$) and gate~2 ($V_{G2}$) at a temperature of
\mbox{$T=290$\,mK} and \mbox{$V_{sd}=-128$\,$\mu$V}. The resulting
honeycomb pattern represents the charge stability diagram of
coupled double quantum dots. Two triple points are marked by blue
dots for clarity. Dashed lines are guides to the eye. (b) Close-up
of a single honeycomb cell. (c) Vicinity of the triple points at a
source-drain bias voltage of \mbox{$391$\,$\mu$V}. (d) and (e)
Capacitive and molecular model of a double quantum dot,
respectively.}
\end{figure}

A magnified colorscale plot of the differential conductance
$dI/dV_{sd}$ in a reduced gate-voltage range is shown in
Fig.~\ref{figure2}(a). The visible high-conductance ridges define
a charge-stability map that is shaped like a honeycomb. This
honeycomb pattern is characteristic of a double quantum dot.
Within each cell, the number of holes (n,m) on the two dots is
constant. Energizing gate~1~(2) to more negative voltages
successively fills holes into dot~1~(2), whereas a more positive
voltage pushes holes out of the dot. Two identical devices were
measured at low temperatures and both exhibited a similar
honeycomb pattern

Of particular importance for sequential tunneling through the
double dot are the so-called triple points, the two blue points in
Fig.~\ref{figure2}(a), for example. At these points, three charge
states are simultaneously degenerate (e.g. $(n,m)$, $(n+1,m)$, and
$(n,m+1)$), enabling the shuttling of a single electron from
source to drain through the two dots. The conductivity in the
vicinity of a triple point strongly depends on the relative
magnitude of the electrostatic and tunnel coupling. For purely
electrostatic coupling, the triple points are sharply defined,
while they become blurred, leading to curved edges, if
quantum-mechanical tunneling is turned on.

We will first analyze the honeycomb pattern, assuming purely
electrostatic interaction as illustrated in Fig.~\ref{figure2}(d).
Hence, we disregard the tunnel coupling between the dots for the
moment. In the quantitative determination of the dot and gate
capacitances, we follow the work of van der Wiel {\it et al}
\cite{VanderWiel}. From the dimensions of a single cell $\Delta
V_{G1,2}= \mid e \mid /C_{G1,2}$ as illustrated in
Fig.~\ref{figure2}(b), one obtains the gate capacitances
\mbox{$C_{G1}=23$\,aF} and \mbox{$C_{G2}=21$\,aF}. Applying a
finite source-drain bias voltage $V_{sd}$ results in a broadening
of the triple points at the honeycomb edges into triangular-shaped
regions, see Fig.~\ref{figure2}(c). In our device the triangles
are less clearly defined due to finite temperature and the strong
tunnel coupling between the dots which we will discuss in the
following paragraphs. Using the relation $C_{G1,2}/C_{1,2} =
|V_{sd}|/\delta V_{G1,2}$, the capacitances $C_{1}=C_S+C_{G1}+C_m$
and $C_{2}=C_D+C_{G2}+C_m$ follow to be \mbox{$84$\,aF} and
\mbox{$145$\,aF}, respectively, from which we obtain $U_{C_{1,2}}
= e^2/C_{1,2}$ $\approx 1.9$\,meV and $1.1$\,meV for the on-site
charging energies of the dots, in agreement with the dimensions of
the Coulomb blockade diamonds at finite bias (not shown). The
mutual capacitance $C_m$ between the two dots can now be estimated
from the triple-point spacing $\Delta V_{G1,2}^{m}$ in
Fig.~\ref{figure2}(b) using $\Delta V_{G1,2}^{m} = |e| C_{m} /
C_{G1,2} C_{2,1}$. We obtain \mbox{$C_{m}\approx 15$\,aF}.

\begin{figure}[t]
\includegraphics*[width=\linewidth]{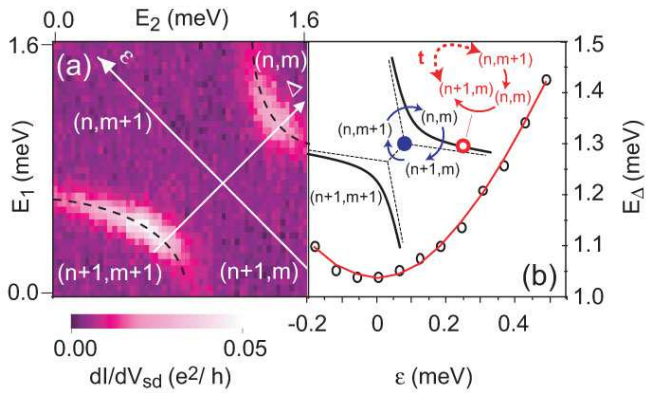}
\caption{\label{figure3}
    (a) Colorscale plot of the differential conductance ($V_{sd}=20$\,$\mu$V, $T = 290$\,mK) in the vicinity of
    two triple points. Dashed lines are guides to the eye.
    (b) Spacing $E_{\Delta}$ (see Eq.~(\ref{eq:EDelta}))
    of the two high conductance wings with respect to
    the $\Delta$-direction versus detuning $\epsilon$.
    Inset:
    Schematics of sequential tunnel processes allowed
    at the triple points (blue dot) and at the honeycomb edges (red circle) via molecular states.}
\end{figure}

We emphasize that disregarding tunneling between the dots
is a very strong assumption.  The purely electrostatic model,
which we have used up to now, overestimates $C_m$ and can only yield an
upper bound. That tunneling is appreciable in this double-dot
system is evidenced by the honeycomb borders in Fig.~\ref{figure2}(a),
which are bright over an extended range. In addition, the high-conductance ridges
are {\em curved} in the vicinity of the triple points, as expected for
strongly tunnel-coupled dots.
Analyzing this curvature allows us to precisely extract the tunnel
coupling amplitude $t$ (see Fig.~\ref{figure3}). To do so, a
convenient description is developed first.

 We adopt a model Hamiltonian of the form $H=H_{C}+H_{T}+H_{L}$,
describing the system depicted in Fig.~\ref{figure2}(e). Here,
$H_{C}$ describes the orbital and Coulomb energies of the
double-dot system,
$H_{T}=t\left(\ket{n+1,m}\bra{n,m+1}+\mathrm{h.c.}\right)$ the
tunnel-coupling between the two dots, and $H_{L}$ the coupling of
each dot to the leads. In $H_C$, we include on-site ($U$) and
nearest-neighbor ($U^{\prime}$) charging energies. States with a
fixed number of charges on each dot are eigenstates of $H_{C}$:
$H_{C}\ket{n,m}=E_{nm}\ket{n,m}$, where
$E_{nm}=E_{nm}^{orb}+\frac{U}{2}\left[n(n-1)+m(m-1)\right]+U^{\prime}nm+E_{1}n+E_{2}m$.
$E_{nm}^{orb}$ is the total orbital energy of the $\ket{n,m}$
charge configuration, and $E_{1(2)}$ is the single-particle energy
of the left (right) dot, supplied by the gate voltages $V_{G1,2}$.
In a simple picture of sequential tunneling~\cite{seq, golovach}
through $H_{C}$-eigenstates (neglecting $H_{T}$ to leading order),
one would expect nonzero conductance only at the triple points. It
is only at these points that energy-conserving processes of the
kind
$\ket{n,m}\rightarrow\ket{n+1,m}\rightarrow\ket{n,m+1}\rightarrow\ket{n,m}$
can lead to charge transport through the double dot (blue sequence
in the inset of Fig.~\ref{figure3}(b)).

However, if we allow for superposed double-dot states of the form
$\ket{E}=\alpha\ket{n+1,m}+\beta\ket{n,m+1}$, sequential transport
is possible along the honeycomb edges as well (red sequence in the
inset of Fig.~\ref{figure3}(b)). Such superposed states are
eigenstates of the full double-dot Hamiltonian $H_C+H_T$. For
spinless holes~\cite{spin} and assuming that only a single
eigenstate $\ket{E}$ participates in transport, the stationary
sequential-tunneling current is then given by
\begin{equation}
  I=\left|e\right|\Gamma [f_{s}(\mu_{2dot})
  -f_{d}(\mu_{2dot})].\label{eq:Current}
\end{equation}
Here,
$f_{l}(\mu_{2dot})=1/\left(\exp\left[(\mu_{2dot}-\mu_{l})/kT\right]+1\right)$
is a Fermi function at temperature $T$, $\mu_{l}$ ($l=s(d)$) the
chemical potential of the source (drain) lead, and
$\Gamma=|\alpha\beta|^{2}\Gamma_s \Gamma_d/(\alpha^2 \Gamma_s +
\beta^2 \Gamma_d)$, with $\Gamma_{s(d)}$ the dot-lead tunneling
rate to the source (drain). The chemical potential of the double
dot $\mu_{2dot}$ depends on whether sequential tunneling occurs at
$\ket{n,m}\leftrightarrow \ket{E}$ (right branch in the inset of
Fig.~\ref{figure3}(b)), or at $\ket{n+1,m+1}\leftrightarrow
\ket{E}$ (left branch): $\mu_{2dot} = E-E_{nm}$ for the former and
$E_{n+1,m+1}-E$ for the latter.

\begin{figure}[t]
\scalebox{0.9}{\includegraphics*[width=\linewidth]{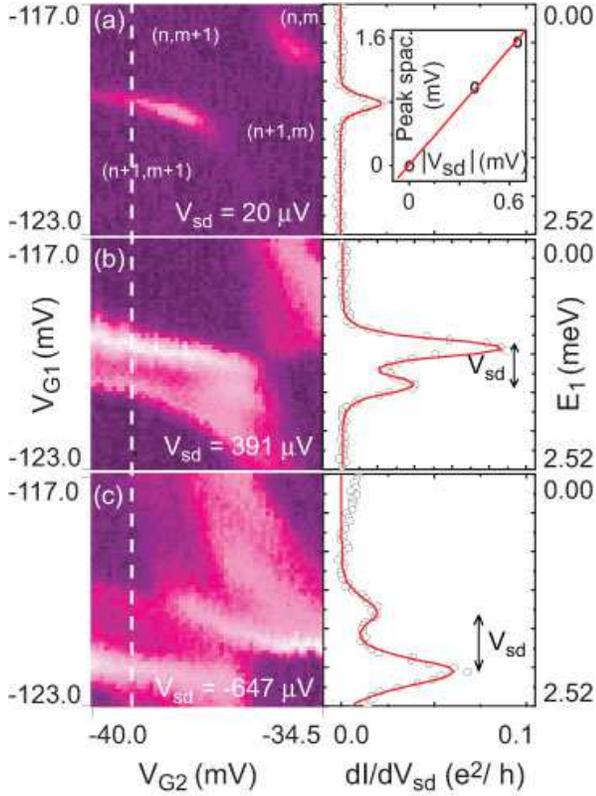}}
\caption{\label{figure4}
    Colorscale plot of the differential conductance in the vicinity of the same triple point as in
    Fig.~3 for three different bias voltage: (a) $V_{sd}=20$\,$\mu$V, (b) $V_{sd}=391$\,$\mu$V and
    (c) $V_{sd}=-647$\,$\mu$V.
    Dark corresponds to $0~e^2/h$ and bright to $0.1~e^2/h$.
    On the right side, open circles denote traces of the differential conductance taken  at the position
    of the dashed line.
    Solid lines represent fits to the line shape given by Eq.~(\ref{eq:DiffCond}).
    Left-hand vertical scale: Voltage applied to gate~1. Right-hand vertical scale:
    Voltage applied to gate~1 converted into energy.}
\end{figure}

With the help of Eq.~(\ref{eq:Current}), the data allow for a
precise quantitative analysis of the tunnel coupling $t$ between
the dots. Figure~\ref{figure3}(a) shows a colorscale plot (linear
scale) of the differential conductance at
$V_{sd}=20\,\mu$V$\approx kT$ in the vicinity of a triple point
region. As expected in the presence of tunnel-coupled eigenstates,
transport is possible not only at the triple points, but also on
the wings extending from the triple points. The two gate voltages
$V_{G1}$ and $V_{G2}$ are converted into energies $E_{1}$ and
$E_{2}$ by multiplying them with the conversion factors
$\alpha_{1}=0.42e$ and $\alpha_{2}=0.29e$, which we obtain from
the splitting of a differential conductance resonance at finite
bias voltage, as will be discussed in the context of
Fig.~\ref{figure4}. We then change variables to
$\epsilon=(E_{1}-E_{2})/\sqrt{2}$ and
$\Delta=(E_{1}+E_{2})/\sqrt{2}$. In terms of these new variables,
the double-dot molecular eigenenergies are (up to a constant
offset)
$E^{\pm}(\Delta,\epsilon)=E_{mn}(\Delta,\epsilon)+\left(\Delta\mp\sqrt{\epsilon^{2}+2t^{2}}\right)/\sqrt{2}$.
When the bias and temperature are smaller than the double-dot
level spacing (i.e., $V_{sd}, kT < E^--E^+$), transport occurs
only through the ground-state $\ket{E^{+}}$. For small bias, we
set $\mu_1=\mu_2=\mu$, then transport is due to energy-conserving
transitions between the state $\ket{E^{+}}$ and either $\ket{n,m}$
(when $E^+-E_{nm}=\mu$) or $\ket{n+1,m+1}$ (when
$E_{n+1,m+1}-E^+=\mu$). These conditions are fulfilled at the two
high-conductance wings. The separation of the wings in the
$\Delta$-direction $(E_\Delta)$ is given by:
\begin{equation}
E_\Delta = \sqrt{2}U^\prime+\sqrt{4\epsilon^2+8t^2}.
\label{eq:EDelta}
\end{equation}
 In Fig.~\ref{figure3}(b) the spacing of the two wings $E_{\Delta}$ is plotted versus the
detuning~$\epsilon$ and fit to Eq.~(\ref{eq:EDelta}). Satisfactory
fits to the data yield a tunnel coupling of $t=310 \dots
360$\,$\mu$eV and $U^{\prime}<100\,\mu eV$. The parameters of the
fit shown are \mbox{$t=358$\,$\mu$eV} and
\mbox{$U^{\prime}=16$\,$\mu eV$}. The relative magnitudes are
compared as $2t \approx 0.7$\,meV $\gg$ $U^{\prime}< 0.1$\,meV.
The fact that the tunnel coupling dominates by almost an order of
magnitude over the electrostatic coupling between the dots
reflects the one-dimensional geometry of a nanotube; electrostatic
interactions are reduced due to the large separation of the
"center of mass" of the charges (while still allowing a
significant overlap of the wavefunctions). Similar molecular
states have been analyzed in semiconductor vertical-lateral double
dots, yielding a smaller tunnel coupling $t \approx 80\mu$eV and
larger U$^{\prime} \approx 175\mu$eV~\cite{Hatano}. Using
\mbox{$U^{\prime}<100$\,$\mu$eV} and
$U^{\prime}=\frac{2e^{2}C_{m}}{C_{1}C_{2}-C_{m}^{2}}$
\cite{Bruder}, one obtains a mutual capacitance of $C_{m}\alt
4\,\mathrm{aF}$, consistent with the previous estimate $C_{m}\le
15\,\mathrm{aF}$ from the purely electrostatic model.

Because $t \gg kT$ at \mbox{$T=0.3$\,K}, charge transport in the
vicinity of the triple points takes place through a single
molecular orbital (the bonding orbital of the two dots). This can
be distinguished from two-stage hopping if $dI/dV_{sd}$ is further
analyzed as a function of bias voltage. More specifically, we
demonstrate next that the finite-bias differential conductance
through the double dot is accurately described by the sequential
tunneling through a single molecular state according to
Eq.~(\ref{eq:Current}).

Figure~\ref{figure4} shows a map of the differential conductance
in the vicinity of the two triple points (same region as
Fig.~\ref{figure3}) for three different source-drain voltages. On
the right side, traces of the differential conductance with
respect to gate~1 are extracted for fixed voltage applied to
gate~2 (dashed line), well separated from the triple points. In
Fig.~\ref{figure4}(a) the conductance trace has a single peak. In
the finite-bias cases (b) and (c) the single peak splits into two
peaks. Because of the linear dependence of the peak splitting on
bias (inset of Fig.~\ref{figure4}(a) for gate~1), the second peak
is not due to an additional level entering the bias window. To
understand this feature, we note that the differential conductance
is measured by modulating the source voltage $\mu_{1}$, keeping
the drain voltage $\mu_{2}$ and all other gate voltages fixed.
Assuming the double-dot charge is fixed, capacitive coupling of
the source to the double dot induces a simultaneous modulation of
$\mu_{2dot}$, albeit with an amplitude reduced by the factor
$r=\partial \mu_{2dot} /
\partial \mu_1 = C_{S}/C_{\Sigma}$, where $C_{\Sigma}\approx
C_{S}+C_{D}+C_{G1}+C_{G2}$. From Eq.~(\ref{eq:Current}) the
differential conductance for our setup is then given by
\begin{equation}
\frac{dI}{d\mu_{1}}=-\left|e| \Gamma
[\left(1-r\right)f_{s}^{\prime}(\mu_{2dot})
+rf_{d}^{\prime}(\mu_{2dot})\right],\label{eq:DiffCond}\end{equation}
where $f_{l}^{\prime}(x)=\frac{d}{dx}f_{l}(x)$. Sequential
tunneling through a single molecular level therefore predicts a
double-peaked structure with \textit{peaks separated by the bias
voltage}, as observed in Fig.~4. The spacing of the two peaks can
thus be used to convert top-gate voltages into energy and one
obtains the conversion factors given above. For our device, we
have $C_{S}\approx 65\,\mathrm{aF}$, $C_{\Sigma} \approx
230\,\mathrm{aF}$, which yields $r\approx 0.3$. According to this
model the relative height of the two differential conductance
peaks should be roughly $\frac{r}{1-r}\approx 0.5$. This value is
consistent with the data shown in Fig.~4 (with ratios of $0.42$ in
(b) and $0.28$ in (c)).
Additionally, we find that the asymmetry of the peaks
switches from positive (b) to negative (c) bias, as is expected
from Eq.~(\ref{eq:DiffCond}).

The data in Fig.~4(a) have been fit to Eq.~(\ref{eq:DiffCond})
yielding a peak width of \mbox{$49$\,$\mu$eV}. Note that in this
case $V_{sd}\approx kT$ and the peak thus does not split.
Subtracting the bias of \mbox{$20$\,$\mu$eV} one obtains an
effective temperature of the electrons of \mbox{$29$\,$\mu$eV}
\mbox{$\approx 335$\,mK}. Fitting Fig.~4(b) and 4(c) to Eq.
(\ref{eq:DiffCond}), one obtains a larger peak width corresponding
to temperatures of \mbox{$785$\,mK} and \mbox{$1180$\,mK},
respectively, which we attribute to Joule heating at finite bias.

The excellent agreement of the sequential-tunneling fits
demonstrates that transport occurs through a single level. In this
regime of a strongly tunnel-coupled double dot, transport cannot
be captured by dot-to-dot hopping, but takes root in the formation
of coherent molecular states.



\begin{acknowledgments}
For assistance and discussions we thank W. Belzig, E. Bieri, C.
Bruder, A. Eichler, V.\ N. Golovach, L. Gr\"uter, G. Gunnarson, D.
Keller, T. Kontos, S. Sahoo, and J. Gobrecht for oxidized
Si-substrates. We acknowledge financial support from the Swiss
NFS, the NCCR on Nanoscience, DARPA (WAC and DL), NSERC (WAC), and
the `C. und H. Dreyfus Stipendium'~(MRG).
\end{acknowledgments}


\bibliographystyle{apsrev}

\begin{thebibliography}{33}
\bibitem{seq}
    We define a sequential-tunneling process for the double dot as a
    process that changes the total charge on the double-dot by
    one, as in Ref.~\cite{golovach}.
\bibitem{spin}
  {Note that in a spinless description we exclude the possibility of e.g. spin-blockade, which, in our
  experiment, has not been observed.}
\end{thebibliography}


\end{document}